\documentclass[twocolumn,aps,pre,showpacs,floatfix]{revtex4}
\usepackage{epsfig}
\usepackage{amsmath}
\usepackage{psfig}
\usepackage{amssymb}
\newcommand{\dd}{\text{d}}

\newcommand{\ee}{\text{e}}

\begin{document}
\title{Magnetization distribution in the transverse Ising chain with energy flux}
\author{V. Eisler$^{1}$, Z. R\'acz$^{1,2}$ and F. van Wijland$^{2}$}

\affiliation{${}^1$Institute for Theoretical Physics,
E\"otv\"os University, 1117 Budapest, P\'azm\'any s\'et\'any 1/a, Hungary\\
${}^2$Laboratoire de Physique Th\'eorique, B\^at. 210,
Universit\'e de Paris-Sud, 91405 Orsay Cedex, France}

\date{January 23, 2003}

\begin{abstract}
The zero-temperature transverse Ising chain carrying an
energy flux $j_E$ is studied
with the aim of determining the nonequilibrium distribution functions,
$P(M_z)$ and $P(M_x)$, of its transverse and
longitudinal magnetizations, respectively.
An exact calculation reveals that $P(M_z)$ is a Gaussian both at
$j_E=0$ and $j_E\not=0$, and
the width of the distribution decreases with increasing energy flux.
The distribution of the order-parameter fluctuations, $P(M_x)$, is
evaluated numerically
for spin-chains of up to 20 spins. For the equilibrium case ($j_E=0$),
we find the expected
Gaussian fluctuations away from the critical point while the critical order-parameter
fluctuations are shown to be non-gaussian with a scaling function
$\Phi(x)=\Phi (M_x/\langle M_x \rangle )=\langle M_x \rangle P(M_x)$
strongly dependent on the boundary conditions.
When $j_E\not=0$, the system displays long-range, oscillating correlations but
$P(M_x)$ is a Gaussian nevertheless, and the width of the
Gaussian decreases with increasing $j_E$. In particular, we find
that, at critical transverse field, the width has a $j_E^{-3/8}$ asymptotic in
the $j_E\to 0$ limit.
\end{abstract}
\maketitle

\section{Introduction}

Nonequilibrium steady states (NESS) have been much studied but a description
of some generality has not emerged so far. Among the many approaches
tried, there are two which continue to receive particular attention.
One of them is an attempt to understand the general features of
NESS through studies of nonequilibrium phase transitions
\cite{{SchmZia},{Marrobook},{Racz-LesH}}.
The basic assumption here is that the universality displayed by equilibrium phase transitions
carries over to critical phenomena in NESS, as well. Thus, by investigating the
similarities and differences from equilibrium, one may gain an understanding of the
role of various components of the competing dynamics generating the steady state.
For example, one may find by observing the universality classes of various nonequilibrium
phase transitions that dynamical anisotropies often yield dipole-like effective
interactions \cite{{Schm},{BZ},{uwerev}}
or that competing non-local dynamics (anomalous diffusion) generates
long-range, power-law interactions \cite{DRT}.

The second approach is more ambitious but less well founded. It concerns the building
(approximating, guessing) of the entropy or the distribution function for
the nonequilibrium microstates \cite{{Tsallis1},{Tsallis2},{Beck}}. The
idea here is that NESS is often the result of complex nonlinear dynamics
which yield distributions that cannot be described by the Boltzmann-Gibbs statistics.
The reason e.g. may be that the dynamics generates long-range
correlations which results in the loss of additivity of such quantities as
the entropy. Furthermore, the dynamics may concentrate the
distribution onto a fractal subspace in the long-time limit, a case that
again may be problematic in equilibrium statistics. In order
to treat such cases, the so called non-extensive statistical
mechanics was introduced \cite{Tsallis1}. It is an approach that takes its name
from the non-extensive character of the postulated entropy.
It has been much developed \cite{Tsallis2} as well as criticized
(see e.g. \cite{{Gross},{Nauenberg}} for the latest examples)
during the last decade. Not surprisingly, the approach had its success
in connection with systems which
have long-range interactions or display (multi)fractal behavior \cite{Tsallis2}.

Our work presented below is perhaps best viewed as an attempt
to build a bridge between the above approaches. On the one hand,
we study the effects of a nonequilibrium constraint on
a well investigated phase transition. Namely, we take the transverse Ising chain
which has an order-disorder transition as the transverse field ($h$) is varied, and
drive it by a field to produce an energy flux, $j_E$ through the system.
The resulting steady states have been investigated \cite{ARS97} and it has been found
that, in addition to the equilibrium phases, a flux-carrying nonequilibrium phase
appears which is distinct by its correlations decaying with distance as a power law.
On the other hand, we shall be concerned with the distribution function in the
various phases of the above system. More precisely, we shall determine
the steady-state distribution functions, $P(M_z)$ and $P(M_x)$
of the $M_z$ (non-ordering field) and $M_x$ (ordering field) components of the
macroscopic magnetization in all three phases of the system and at and near
its critical point.

The results are surprisingly simple. The distribution functions are Gaussian in the
equilibrium phases away from the critical point. This is expected since we have
macroscopic quantity and the correlations decay exponentially. The distribution of
the non-ordering field remains a Gaussian at the critical point of the equilibrium
system, as well. The reason for this is that although the appropriate correlations
decay with distance $n$ as a power law but the exponent in the power is large ($1/n^2$),
so that the fluctuations
$\langle M_z^2\rangle - \langle M_z\rangle^2$ do not diverge at $h=h_c$. The
distribution function of the ordering field becomes nontrivial at $h_c$ and
our numerical calculations demonstrate that $P(M_x)$ depends strongly on the
boundary conditions taken to be periodic, anti-periodic, and free. The unexpected
simplicity is in the current-carrying phase where the energy flux
generates long-range correlations decaying as a power law ($1/\sqrt{n}$)
but, nevertheless, the distributions are gaussian.
The mathematical reason for this lies in the oscillating character
of the correlations which prevents the divergence of the
spatial sum of the correlations which in turn are proportional to the fluctuations.
Physically, the oscillations in the correlations can be traced to the form of
energy flux [see eq.(\ref{eflux}) below] which suggest that the consecutive
$x$ and $y$ components of the spins are more and more rigidly interconnected
as $j_E$ is increased and thus fluctuations decrease with increasing $j_E$.
This picture will be seen to be valid near the nonequilibrium phase
boundaries where the fluctuations as a function of $j_E$ can be explicitely
calculated.

The above results are presented in the following order. Section \ref{TRIjE} contains a
review of the transverse Ising model with energy flux, including the setup of the
formalism convenient for calculating the distribution functions. Next (Sec.\ref{Pzdist}),
the distribution $P(M_z)$ is calculated exactly. Numerical work on $P(M_x)$
and preliminary analytic work on correlations are presented
in Sec.\ref{NUM}, followed by concluding remarks in
Sec.\ref{Final}.
\section{Transverse Ising model with energy flux}
\label{TRIjE}

The transverse Ising chain is one of the simplest system displaying
a critical order-disorder transition \cite{LSM}. It is defined by the Hamiltonian
\begin{equation}
\hat{H_I}=-J\sum_i s_i^x s_{i+1}^x-\frac{h}{2}\sum_i s_i^z
\label{TRI}
\end{equation}
where $\vec{s}_i=\frac{1}{2}\vec{\sigma}_i$ and $\sigma_i^\alpha$ ($\alpha=x,y,z$)
denotes the three Pauli matrices at sites
$i=1,2,...,N$ of a $d=1$ periodic chain
($s^\alpha_{N+1}=s^\alpha_1$), and $h$ is the
transverse field in units of the Ising coupling ($J=1$ is set in the rest of the paper).

The order parameter of this system is $M_x=\sum_i s_i^x$, and the ground state
of this Hamiltonian changes from being disordered $\langle M_x\rangle =0$ for $h>1$
to ordered $\langle M_x\rangle \not= 0$ for $h<1$. The transition point $h=h_c$ is a
critical point in the universality class of the two-dimensional Ising model.

\begin{figure}[htb]
  \centerline{ \epsfxsize=8cm \epsfbox{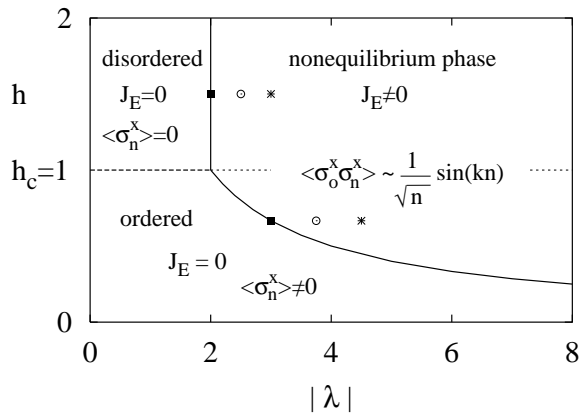} }
   \vspace{0.cm}
\caption{Phase diagram of the driven transverse Ising model
        in the $h-|\lambda |$ plane where $h$ is the transverse field while
        $\lambda$ is the effective field which drives the flux of energy.
        Pairs of dual-conjugate points are shown by filled squares, circles,
        and stars; and the line $h=1$ is self-dual as discussed in the text.
        Power-law correlations are present in the nonequilibrium phase,
        $J_E\not= 0$, and on the Ising critical line in the
        equilibrium phase, $J_E=0$ ($h=1$, $0\le |\lambda |\le 2$).}
\label{fig:phase-diag}
\end{figure}

In order to drive the above system out of equilibrium, one can add a bulk field
that drives the energy flux. The simplest bulk field is the energy flux itself
\cite{ARS97} thus, in order to produce a nonequilibrium state of $\hat H_I$ one
should find the ground state of the following Hamiltonian
\begin{equation}
\hat{H}=-\sum_i s_i^x s_{i+1}^x-\frac{h}{2}\sum_i s_i^z-\lambda\hat{J_E}\, .
\label{fullham}
\end{equation}
Here the driving field $\lambda$ is again measured in units of $J$, and
the current operator $\hat{J}$ is the sum of local energy flux given by the
following expression
\begin{equation}
\hat{J_E}=\frac{h}{4}\sum_i(s^x_i s^y_{i+1}-s^y_i s_{i+1}^x) \, .
\label{eflux}
\end{equation}
The driven system defined by (\ref{fullham}) and (\ref{eflux}) can be solved
\cite{ARS97} and one finds that the ground state does not change and the
energy flux is zero up to a critical value
$\lambda=\lambda_c(h)$ of the driving field. The ground state expectation value
of the energy flux
$j_E=\langle \hat J_E\rangle/N$ becomes nonzero for $|\lambda |>\lambda_c$.
The resulting phase diagram is depicted in Fig.\ref{fig:phase-diag}.

Here we note a hitherto unnoticed property of the Hamiltonian (\ref{fullham}).
Namely, the duality properties of the transverse Ising model \cite{duality}
have an appropriate generalization to the full nonequilibrium phase
diagram of $\hat H$. Indeed, let us denote
the action of duality transformation by $s_i^\alpha\to s_i^{\alpha\star}$
and define the transformation as it is done for
the Ising model in a transverse field
\begin{equation}
s_i^{z\star}=2s_i^x s_{i+1}^x,\;s_i^{x\star} s_{i+1}^{x\star}=\frac 12
s_i^z \, .
\label{duality}
\end{equation}
It can be verified then that the Hamiltonian
$\hat{H}[h,\lambda,\{s_i^\alpha\}]$ (\ref{fullham}) characterized by the two couplings
$(h,\lambda)$ transforms into an identical Hamiltonian with couplings
$(h,\lambda)^\star=(h^{-1},-\lambda h)$
\begin{equation}
\hat{H}[h,\lambda,\{s_i^\alpha\}]=h\hat{H}[h^{-1},-\lambda
h,\{s_i^{\alpha\star}\}] \, .
\end{equation}
Hence the duality transformation leaves the whole $|\lambda |=\lambda_c$ curve
globally invariant and, furthermore, it leaves the $h=1$ line pointwise invariant
(examples of dual-conjugate points are given in Fig.\ref{fig:phase-diag}). In order to
keep the formulas simple, from
now on we shall restrict our analysis
to $\lambda\ge 0$, that is to $j_E\ge0$.

The self-dual $h=1$ line is expected to display special properties. For example,
quantities such as $\lambda_c/\lambda$, or the wavevectors $q_{\pm}$
where the excitation spectrum is gapless, are left invariant by the duality
transformation. Furthermore, the functional form of various physical quantities
(dispersion, energy flux, fluctuations) considerably simplify on this
line and thus the knowledge of selfduality helps
locating limits where exact calculation can be carried out.

Our main goal is to calculate the distribution functions $P(M_z)$ and $P(M_x)$
in various regions in the above phase diagram. These functions are
defined as
\begin{equation}
P(M_\alpha)=\langle \delta(M_\alpha-\sum_i{s_i^\alpha})\rangle
\label{genP}
\end{equation}
where brackets denote
the ground-state expectation value (note that we have omitted
the index $\alpha$ in $P_\alpha(.)$, i.e. the argument of the
function defines which distribution is considered).

There are two parts to our calculations. The function $P(M_x)$ is
evaluated numerically by diagonalizing $\hat H$ for chains
containing up to $N=18-20$ spins, while $P(M_z)$ is found
analytically. The exact calculation is possible because $M_z=\sum_q (c_q^+c_q -1/2)$
is a quadratic form in the
fermion operators ($c_q^+$, $c_q$) in which the Hamiltonian
$\hat H=\hat H_I-\lambda \hat J_E$ is quadratic as well. Thus the calculation of
the generating function
\begin{equation}
G(s)=\langle \ee^{-s M_z}\rangle
\label{genfunc}
\end{equation}
becomes a problem of evaluating Gaussian integrals. We begin with this part
of the problem.

\subsection{Formalism}

The Hamiltonian $\hat H$ can be diagonalized \cite{ARS97}
by first introducing
creation-annihilation operators, then employing
the Jordan-Wigner transformation \cite{LSM} to transform them
into fermion operators
($c_\ell, c_\ell^+$) and,
finally, using a Bogoljubov transformation
on the $c_q$ and $c_{-q}^+$ components of the
Fourier transforms of $c_\ell$-s.
The calculation of $P(M_z)$ becomes relatively simple if after
using the Jordan-Wigner transformation one passes to a path-integral
formulation (see {\it e.g.}
\cite{negeleorland} for a
pedagogical account). One then finds that the system
is described by the following quadratic action
\begin{eqnarray}
S[\bar{c},c]=\quad &\nonumber \\ \nonumber \\
\int\frac{\text{d} \omega}{2\pi}\int\frac{\text{d}
q}{2\pi}&\left[
\frac{1}{2}(\bar{c}_q(\omega)\;c_{-q}(-\omega))B_{q,\omega}\left(
\begin{array}{c}\bar{c}_{-q}(-\omega)\\c_{q}(\omega)\end{array}
\right)\right]
\end{eqnarray}
where the Grassman fields $\bar{c}_q(\omega)$ and $c_{q}(\omega)$ are related
to $c_q^+$ and $c_{q}$ correspondingly, while the scattering
matrix $B_{q,\omega}$ has the following inverse
\begin{equation}
B_{q,\omega}^{-1}=\left(\begin{array}{cc}
\langle \bar{c}_{-q}(-\omega)\bar{c}_q(\omega)\rangle&
\langle \bar{c}_{-q}(-\omega){c}_{-q}(-\omega)\rangle\\
\langle{c}_{q}(\omega)\bar{c}_q(\omega)\rangle&
\langle c_q (\omega)c_{-q}(-\omega)\rangle\end{array}\right) \, .
\end{equation}
Here the correlators are given by
\begin{eqnarray}
\nonumber
\langle{c}_{q}(\omega)\bar{c}_q(\omega)\rangle=
\langle{c}_{q}(\omega)\bar{c}_q(\omega)\rangle^*=  \hspace{2.8cm}\\
\nonumber \\
\frac{1}{2\Lambda_q}\left[\frac{\frac{h}{2}+\frac{1}{2}\cos
q-\Lambda_q}{i\omega-\Lambda_q^+}-\frac{\frac{h}{2}+
\frac{1}{2}\cos q+\Lambda_q}{i\omega-\Lambda_q^-}\right]
\end{eqnarray}
and
\begin{eqnarray}\nonumber
\langle \bar{c}_{-q}(-\omega)\bar{c}_q(\omega)\rangle=
\langle {c}_{q}(\omega){c}_{-q}(-\omega)\rangle^*=\hspace{1.8cm}\\ \nonumber\\
\frac{i\frac{1}{2}\sin q}{2\Lambda_q}
\left(\frac{1}{i\omega-\Lambda^{-}_q}-\frac{1}{i\omega-\Lambda_q^{+}}\right)
\end{eqnarray}
where $\Lambda_q$ and $\Lambda_q^\pm$ are the dispersion relations for
$\hat H_I$ and $\hat H$, respectively
\begin{eqnarray}
\Lambda_q=\frac 12 \sqrt{1+h^2+2h\cos q} \\
\Lambda^{\pm}_q=\pm\Lambda_q+\frac{\lambda h}{4}\sin q \, .
\label{dispersion}
\end{eqnarray}
In order to return to the time variables,
we must first study the two branches $\Lambda^\pm_q$ of the spectrum.

\subsection{Spectrum and energy flux}

Flux of energy is present in the system only above a
critical drive $\lambda>\lambda_c$
\cite{ARS97} (see Fig.\ref{fig:phase-diag}) where
\begin{equation}
\lambda_c(h)=\left\{\begin{array}{l}2\text{ if }h\geq 1 \\\frac{2}{h}\text{ if
}h\leq 1\end{array}\right. \, .
\end{equation}
Indeed, it is not hard to see that
\begin{equation}
\Lambda^+_q(\lambda\leq \lambda_c)>0\; \; ;
\; \;\Lambda_q^-(\lambda\leq \lambda_c)<0 \, .
\end{equation}
Hence the ground state is unchanged with respect to the pure Ising model ground
state as long as the driving field does not exceed $\lambda_c(h)$.
This means that all observables will assume their Ising model values and no energy
current will be flowing through the chain.

However, for $\lambda\geq \lambda_c$ one may see that $\Lambda^+_q$ (resp.
$\Lambda^-_q$) changes sign over the interval $I_2=[q_-,q_+]$ (resp. $I_4=[-q_+,-q_-]$).
The explicit expression of the wave vectors $q_\pm$ is deduced from
\begin{equation}
\cos q_\pm=\frac{-4\pm\sqrt{(\lambda^2 h^2-4)(\lambda^2-4)}}{\lambda^2
h},\;\;-\pi\leq q_{\pm}\leq 0 \, .
\end{equation}
Beyond the critical drive, the excitation spectrum gives rise to a
ground state which breaks the left-right symmetry
and, indeed, it may be verified \cite{ARS97} that a nonzero energy flux
is present in the chain with the explicit form of the flux is given by
\begin{equation}
j_E=\frac{h}{4\pi\lambda^2}\sqrt{\left(\lambda^2-\frac{4}{h^2}\right)(\lambda^2-4)}
\, .
\label{fluxequation}
\end{equation}
For later applications we specify the $j_E=j_E(\lambda)$ function in
the vicinity of the critical drive, as $\lambda\to\lambda_c^+$, 
\begin{equation}\label{cor-lambda-jE}
j_E\approx\left\{
\begin{array}{rl}
\frac{\sqrt{h^2-1}}{4\pi}\sqrt{\lambda-2}&\text{ for }h>1,\lambda_c=2\\
\frac{1}{4\pi}(\lambda-2)&\text{ for }h=1,\lambda_c=2\\
\frac{h^{3/2}\sqrt{1-h^2}}{4\pi}\sqrt{\lambda-\frac{2}{h}}&\text{ for
}h<1,\lambda_c=\frac{2}{h}\end{array}
\right.
\end{equation}
Besides, as they will naturally arise in the upcoming discussion,
we further define here intervals
$I_1=[-\pi,q_-]$, $I_3=[q_+,-q_+]$ and
$I_5=[-q_-,\pi]$ which are complementary to $I_2$ and $I_4$ in $[-\pi,\pi]$.

\subsection{Inverse of the scattering matrix in the absence or presence of
energy flux}

When no current flows through the system, $j_E=0$, the equal-time transform of
$(B_{q,\omega})^{-1}$, denoted by $B_q^{-1}$, reads:
\begin{equation}\label{inverseB-J=0}
B_q^{-1}=\left(\begin{array}{cc}
\frac{i\frac{1}{2}\sin q}{2\Lambda_q}&\frac{\frac{h}{2}+\frac{1}{2}\cos
q+\Lambda_q}{2\Lambda_q}\\
-\frac{\frac{h}{2}+\frac{1}{2}\cos q+\Lambda_q}{2\Lambda_q}&
-\frac{i\frac{1}{2}\sin q}{2\Lambda_q}\end{array}\right)
\end{equation}

For $j_E\not=0$, on the other hand, $B_q^{-1}$
is given by
Eq.~(\ref{inverseB-J=0}) only for $q\in I_1\cup I_3\cup I_5$.
If $q\in I_2\cup I_4$, its expression is changed to
\begin{equation}\label{inverseB-Jnot=0}
I_2:\;\;B_q^{-1}=
\left(\begin{array}{cc}
0&0\\
-1&0
\end{array}\right),\;\;\;I_4:\;\;B_q^{-1}=\left(\begin{array}{cc}
0&1\\
0&0
\end{array}\right)
\end{equation}
Having the expressions for $B_q^{-1}$, we can start the calculation of the
distribution function $P(M_z)$.

\section{Distribution function for the transverse magnetization}
\label{Pzdist}
\subsection{Calculation of the generating function and its moments}
The generating function (\ref{genfunc}) of $P(M_z)$ can be expressed through
fermionic operators as
\begin{equation}
G(s)=\langle \ee^{-s M_z}\rangle=\text{e}^{N\frac{s}{2}}\langle
\text{e}^{-s\sum_q c^{\dagger}_q c_q}\rangle \, .
\end{equation}
After normal ordering $\text{e}^{-s c^{\dagger}_q c_q}$ and using the Grassman fields,
we are left with evaluating the following expression
\begin{equation}
G(s)=\text{e}^{N\frac{s}{2}}\langle\text{e}^{-\tilde{s}\sum_q \bar{c}_q
c_q}\rangle,\;\;\;\tilde{s}=1-\text{e}^{-s}
\label{gengrass}
\end{equation}
and the fields in the exponentials are evaluated at some fixed time.
In order to evaluate (\ref{gengrass}), we recall
the following result for Grassmann integrals \cite{negeleorland}: the vacuum
expectation value of observables of the form $\text{e}^{-\sum_q
{z}_q\bar{c}_q c_q}$ can be obtained as:
\begin{eqnarray}\nonumber
\langle\text{e}^{-\sum_q{z}_q\bar{c}_q
c_q}\rangle=\qquad\qquad\qquad\qquad\qquad\qquad\qquad\\\prod_q\sqrt{\det
\left(\begin{array}{rr}-\sqrt{{z}_q{z}_{-q}}(B_q^{-1})_{11}
&-1+{z}_{-q}(B_q^{-1})_{12}\\
1+{z}_q(B_q^{-1})_{21}&
-\sqrt{{z}_q{z}_{-q}}(B_q^{-1})_{22}\end{array}\right)
}\label{Grformule}
\end{eqnarray}
As one can see $G(s)$ is a special case of eq.~(\ref{Grformule}) and it
can be evaluated by using the appropriate expressions (\ref{inverseB-J=0})
or (\ref{inverseB-Jnot=0}) for $B_q^{-1}$.

If no current flows, that is for $\lambda\leq \lambda_c$, we find that $\ln G$
(the cumulant generating function) is given by
\begin{equation}
\ln G(s)=\frac{N}{2}\int\limits_{-\pi}^\pi \frac{\text{d}
q}{2\pi}\ln\left[\left(1-n_q\right)\text{e}^{s}+n_q\text{e}^{-s}\right]
\label{distribj=0}
\end{equation}
where
\begin{equation}
n_q=(h+\cos q+2\Lambda_q)/(4\Lambda_q) \, .
\label{nq}
\end{equation}
One can verify that the normalization condition $G(0)=1$ is satisfied, and one can
also recover the well-known result found by Pfeuty~\cite{pfeuty} for the magnetization
\begin{equation}
\left.-\frac{\partial \ln G(s)}{\partial s}\right |_0=\langle M_z\rangle=
N\int_{-\pi}^\pi\frac{\text{d} q}{2\pi}
\left(n_q-\frac{1}{2}\right) \, .
\label{Mzaver}
\end{equation}

As we shall see below, $P(M_z)$ is a Gaussian thus, in addition to
$\langle M_z\rangle$, the variance of the transverse
magnetization
$N\sigma^2(h)=\langle M_z^2\rangle-\langle
M_z\rangle^2$ will characterize the distribution. It can be
obtained from the second derivative of $\ln G(s)$ as
\begin{equation}
\sigma^2(h)=2\int_{-\pi}^\pi\frac{\text{d} q}{2\pi}n_q(1-n_q)
\label{sigmaIsing}
\end{equation}
It is interesting to note that the fluctuations in $M_z$ are independent
of the magnetic field in the ordered phase
\begin{equation}\label{sigmaIsing2}
\sigma^2(h)=\left\{
\begin{array}{l}
{1}/{4}\quad \quad \,\;\text{ for }|h|\leq 1\\
1/(4h^2)\quad \text{for }|h|>1\end{array}
\right.
\end{equation}

In the presence of nonzero energy flux ($\lambda> \lambda_c$), the generating function
is more complicated only because of the limits of integration in (\ref{distribj=0})
\begin{equation}
\ln G_\lambda^{}(s)=\frac{N}{2}\int\limits_{I_1\cup I_3\cup I_5} \frac{\text{d}
q}{2\pi}\ln\left[\left(1-n_q\right)\text{e}^{s}+n_q\text{e}^{-s}\right]
\label{distribjnot=0}
\end{equation}
Accordingly, the first and second cumulants of the magnetization are given by
\begin{equation}
\langle M_z\rangle_\lambda^{}=
\frac{N}{2}\int\limits_{I_1\cup I_3\cup I_5}\frac{\text{d} q}
{2\pi}\frac{h+\cos q}{2\Lambda_q}
\end{equation}
\begin{equation}\label{sigmaCourant}
\sigma^2_\lambda(h)=\frac{1}{2}\int\limits_{I_1\cup I_3\cup I_5}
\frac{\text{d} q}{2\pi}\frac{\sin^2 q}{1+h^2+2h\cos q}
\end{equation}
where we have used eq.(\ref{nq}) to write out the integrands explicitely.
Note that we added a $\lambda$ subscript to $G$, $\langle M_z\rangle$ and $\sigma^2$
in order to indicate that these quantities do depend on $\lambda$ for
$\lambda> \lambda_c$.

\subsection{The transverse magnetization distribution is a Gaussian}

In order to show that the transverse magnetization distribution is a Gaussian,
let us consider the $n^{\text{th}}$ cumulant $\langle M_z^n\rangle_c$
which is the coefficient in front of
$(-s)^n/n!$ in the expansion of $\ln G(s) $. The latter coefficient is
$N$ times an integral of a polynomial in $n_q$, both in the current
carrying and current free phases and, furthermore,
$n_q$ is a nonsingular, strictly positive function of $q$ (note that
$n_q$ is finite even at $h=1$).
Hence each cumulant depends linearly on $N$.
In particular, as we have seen, the variance
of $M_z$ denoted by
$N\sigma^2$  and $\sigma^2$ (in Eq. (\ref{sigmaCourant}) or (\ref{sigmaIsing})) is finite.

\begin{figure}[htb]
  \vspace{-2cm}
  \centerline{ \epsfxsize=6cm \epsfbox{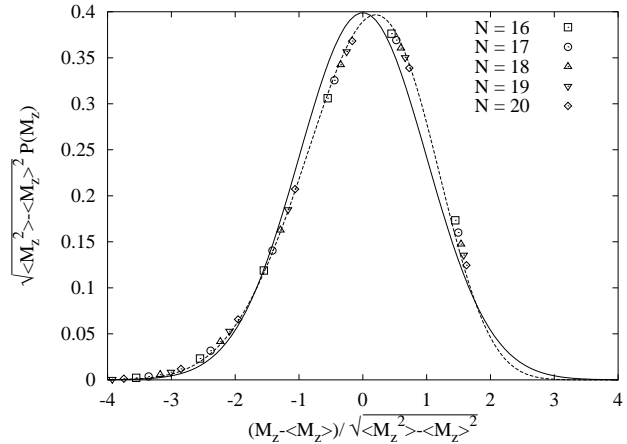} }
   \vspace{0.cm}
\caption{Distribution function $P(M_z)$ for the longitudinal magnetization $M_z$
on the critical line $h=h_c=1$, $\lambda\le 2$. Results for periodic
boundary conditions are displayed. The solid line is the asymptotic Gaussian
while the dashed line is the displacement distribution of a 20-step random walk
of steplength
$1/2$ having a drift generating an average displacement equal to
$\langle M_z \rangle$.}
\label{fig:m_zGauss}
\end{figure}

It follows from the linear $N$ dependence of the cumulants that
\begin{equation}
\frac{\langle M_z^n \rangle_c}{\langle M_z\rangle_c^{n/2} }\sim
\frac{N}{N^{n/2}}\sim N^{(2-n)/2}
\end{equation}
thus the above ratio
goes to zero for all $n>2$ as $N\to \infty$. We may therefore conclude that the limiting
form of the distribution function of the transverse magnetization is a Gaussian
of variance $N\sigma^2$:
\begin{equation}
P(M_z)=\frac{1}{\sqrt{2\pi N \sigma^2}}\ee^{-\frac{(M_z-\langle
M_z\rangle)^2}{2 N \sigma^2}}
\end{equation}
The above result applies over the whole phase diagram and it is useful to
check the numerical procedure employed in Section \ref{NUM}
by evaluating $P(M_z)$ for finite-size systems. As can be seen
in Fig.\ref{fig:m_zGauss}, there is a convergence to the limiting form
with increasing $N$ and, furthermore, the
nearly Gaussian fluctuations of $M_z$ are observed already at small
sizes ($N=16-20$). It is remarkable that the deviations from the Gaussian
are very close to those of an $N$ step random walk with a drift determined
from a correspondence between the left (right) moves and the up (down) spins
generating the average $\langle M_z \rangle$.

\subsection{Width of the Gaussian}

Recalling the expression of the average and the variance
(\ref{Mzaver},\ref{sigmaIsing})
we calculate them in the limit of vanishing flux
($\lambda\to\lambda_c^+$). Let us define
\begin{equation}
\delta M_z=\langle M_z\rangle(h,\lambda)-\langle M_z\rangle(h),\;\;\delta
\sigma^2=\sigma^2(h,\lambda)-\sigma^2(h)
\end{equation}
The above quantities exhibit singular behavior as one enters the current-carrying
phase. For the magnetization we find
\begin{equation}
\delta M_z=-N\int\limits_{q_-}^{q_+}\frac{\dd q}{2\pi}\frac{h+\cos q}{\Lambda_q}
\end{equation}

\begin{equation}
\delta M_z=\left\{
\begin{array}{ll}
\frac{-1}{\pi}(1-h^{-2})(\lambda-\lambda_c)^{1/2}&\text{ for }h>1,\lambda_c=2\\
\frac{-1}{\pi}(\lambda-\lambda_c)&\text{ for }h=1,\lambda_c=2\\
\frac{-1}{\pi}h^{5/2}(\lambda-\lambda_c)^{3/2}&
\text{ for }h<1,\lambda_c=\frac{2}{h}\end{array}\right.
\end{equation}
and for the variance
\begin{equation}\begin{array}{l}
h\geq 1,\lambda_c=2,\;\;\delta \sigma_z^2
\simeq-\frac{1}{\pi h^2}\sqrt{\lambda-\lambda_c}\\
h\leq 1,\lambda_c=\frac{2}{h},\;\;\delta \sigma_z^2
\simeq -\frac{\sqrt{h}}{\pi}\sqrt{\lambda-\lambda_c}
\end{array}\label{deltavar}
\end{equation}
Using (\ref{cor-lambda-jE}) we find, as $j_E\to 0$,
\begin{equation}\label{deltavar2}\begin{array}{l}
h\neq 1,\;\delta \sigma_z^2
\simeq-j_E\\
h=1,\;\delta \sigma_z^2
\simeq -j_E^{1/2}
\end{array}
\end{equation}
As can be seen, the variance of the transverse
magnetization is smaller in the current carrying phase than in the current free
phase. This supports the view that imposing a current stiffens the system,
and thus decreases fluctuations.

\section{Distribution function for the longitudinal magnetization}
\label{NUM}

The exact evaluation of $P(M_x)$ appears to be a nontrivial task and
we have been able to calculate it only numerically for finite size
chains. Since the expression
(\ref{genP}) for $P(M_x)$ is a ground-state expectation value, we had
to find the ground-state
wave function and, due to the sparseness of the Hamiltonian matrix, the
Lanczos algorithm \cite{Lanczos} could be used effectively. Since
the ground state wave function is needed with precision, we were able to accomplish
this task for chain lengths of up to $N=20$ with the results displayed on
Figs. \ref{fig:m_xcrit-per}-\ref{fig:m_xGauss}.

\subsection{Equilibrium distribution}
\label{eqdist}
\begin{figure}[htb]
   \vspace{-2.5cm}
  \centerline{ \epsfxsize=6cm \epsfbox{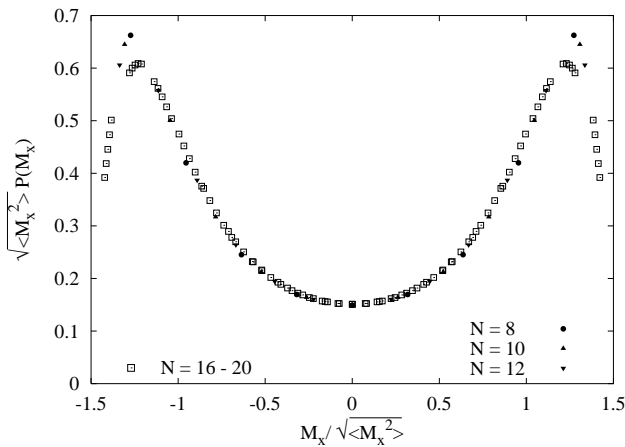} }
   \vspace{0.cm}
\caption{Scaling function for the distribution $P(M_x)$
of the longitudinal magnetization $M_x$
on the critical line $h=h_c=1$, $\lambda\le 2$, for periodic
boundary conditions. In order to demonstrate the smallness of
finite-size effects, the small systems ($N=8,10,12$) are displayed
by full symbols while the larger systems ($N=16-20$) are all
shown by a single empty symbol.}
\label{fig:m_xcrit-per}
\end{figure}

In the equilibrium system ($j_E=0$), the correlation length is infinite
only at the critical point. Thus one expects $P(M_x)$ to be
a Gaussian for $h>1$, a sum of two Gaussians for $h<1$, and a nontrivial
distribution emerges only at criticality ($h=h_c=1$).
This is indeed what we observe, apart from the finite size effects
showing up in nongaussian corrections close to $h=1$.
At the critical point itself, $P(M_x)$ shows fast convergence to the
asymptotic form as can be seen on Fig.\ref{fig:m_xcrit-per} where
the $N\ge 16$ points appear to have settled on the asymptotic curve. This means
that the $N$ dependence of $P(M_x)$ is almost all in the scaling
variable $M_x/\sqrt{\langle M_x^2\rangle}$, a remarkable feature that
has been observed in a series of equilibrium- and nonequilibrium
critical states \cite{{raczdistrib1},{raczdistrib2},{KPZdistr},{PortelliJPA},{BramPRE}}.
Note also that the finite-size effects show up mainly in the large
$M_x/\sqrt{\langle M_x^2\rangle}$
region. This is in accord with the general observation that the
large-argument region of the scaling function is related to the
long-wavelength properties of the system.

\begin{figure}[htb]
   \vspace{-2.5cm}
  \centerline{ \epsfxsize=6cm \epsfbox{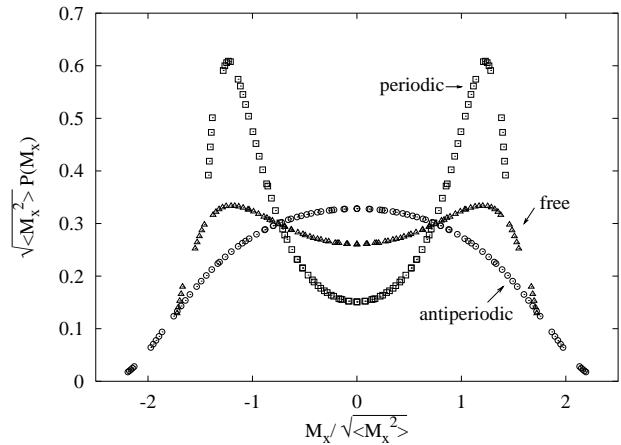} }
   \vspace{0.cm}
\caption{Distribution function $P(M_x)$ for the longitudinal magnetization $M_x$
at the critical line $h=h_c=1$, $\lambda\le 2$. Results for periodic, free, and
antiperiodic boundary conditions are displayed for system sizes $N=16-20$.}
\label{fig:m_xcrit}
\end{figure}

Fig.\ref{fig:m_xcrit} shows the critical point scaling functions
for various boundary conditions (periodic $s^\alpha_{N+1}=s^\alpha_1$,
antiperiodic $s^\alpha_{N+1}=-s^\alpha_1$, and free).
One can observe here
not only the strongly nongaussian character of the distributions but also
the fact that scaling functions do vary with changing the boundary conditions.
The boundary condition dependence of the critical scaling functions is known
\cite{{PrivFish},{Wubook},{Kaneda},{CKHu}}. It is also known that the
scaling functions depend on the shape of the system as well. In case of the
$d=2$ Ising model, this means that the scaling function depends on the
aspect ratio, $a$, of a rectangular sample. Since the transverse Ising model
has its origin in the transfer matrix of the $d=2$ Ising model in an anisotropic
limit \cite{Wubook}, we believe that the distribution functions displayed
in Fig.\ref{fig:m_xcrit} are equal to the $d=2$ critical order parameter distributions
in the $a\to 0$ limit with the boundary conditions in the
"short" direction being the same in the $d=1$ and $d=2$ systems.
Implicit in this belief is the assumption that the boundary conditions in the
"long" direction do not affect the scaling function provided $a\to 0$.

\subsection{Nonequilibrium distribution}
\label{neqdist}

\begin{figure}[htb]
  \vspace{-2.3cm}
  \centerline{ \epsfxsize=6cm \epsfbox{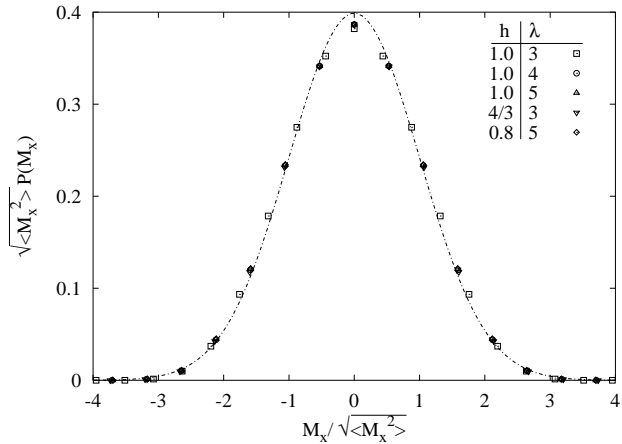} }
   \vspace{0.cm}
\caption{Distribution function $P(M_x)$ for the longitudinal magnetization $M_x$
away from the critical line $h=h_c=1$, $\lambda\le 2$.}
\label{fig:m_xGauss}
\end{figure}

In the nonequilibrium case ($j_E\not=0$), we find that similarly to $M_z$,
the fluctuations of the longitudinal magnetization $M_x$ are also Gaussian
(Fig.\ref{fig:m_xGauss}).
Remarkably, the finite size, non-gaussian corrections
are small even near the $h=1$, $\lambda=2$ point.

The Gaussian result is somewhat
surprising since the flux generates power-law correlations in
$\langle s^x_0s^x_n\rangle$ decaying
with distance as $1/\sqrt{n}$ \cite{ARS97}. Thus one may presume that the
current carrying states are effectively critical. This is not so, however,
since the correlations
are oscillating and the fluctuations $\langle M_x^2\rangle$ (given by the integral of the
correlations) are finite. Actually $\langle M_x^2\rangle$ is decreasing
with increasing $j_E$ (see below) and thus we see here
another example of fluxes generating
power-law correlations but nevertheless making the system more rigid.

The decrease of $\langle M_x^2\rangle$ with increasing $j_E$ can be seen
in the numerical studies of the finite spin chains. On the selfdual
line ($h=1$), however, this can be shown analytically in the limit of
vanishing flux ($\lambda \to \lambda_c=2$) where one finds
\begin{equation}
\langle M_x^2\rangle\propto N j_E^{-3/8} \, .
\end{equation}
The derivation of the above result is possible because the ($h=1$, $\lambda=\lambda_c$)
point is a critical point with infinite correlation length. Approaching this
point along the $h=1$, $\lambda\to\lambda_c^+$ line, one can observe from
numerical studies \cite{ARS97} that the
wavelength $\kappa^{-1}$ of the oscillation of the correlation function
$C_x(r)=\langle s_i^x s_{i+r}^x\rangle\propto\cos{\kappa r}$ diverges
as $\kappa^{-1}\propto (\lambda -\lambda_c)^{-1/2}$. This
diverging wavelength allows one to take a continuum limit and
to establish \cite{preparation} to all orders in perturbation theory, that
the order-parameter correlations possess the following scaling form
\begin{equation}
C_x(r)\approx\frac{{\cal A}}{r^{1/4}}{\cal F}(\kappa r) \, .
\label{corrscaling}
\end{equation}
with the small argument limit of the scaling function explicitely
given by ${\cal F}(x)=1-{x^2}/{2}+{x^4}/{16}+{\cal O}(x^6)$.

The derivation of the above results follows the ideas
\cite{{barouchcoy},{itzyksonbander},{wucoytracybarouch},{wu}}
used for the calculation
of the order-parameter fluctuations in the equilibrium transverse Ising model.
Namely, the correlations are expressed as a Pfaffian of a block  Toeplitz matrix
constructed of $2\times 2$ matrices. Then, in the equilibrium case,
the analysis of the Toeplitz matrices in the
asymptotic limit of $r\to \infty$, and $h\to 1$ with $r(h-1)$ kept fixed
yields $C^{eq}_x(r)\sim {r^{-1/4}}F(r(h-1))$. A similar asymptotic analysis
in the current-carrying phase, using the scaling limit
$r\to \infty$, and $\kappa\sim (\lambda -\lambda_c)^{1/2}\to 0$ with
$\kappa r$ kept finite, results in eq.(\ref{corrscaling}). The derivation is
rather technical and we shall present it, together with the analysis of
other scaling limits, in a separate publication \cite{preparation}.

Once the correlations are known, the fluctuations can be calculated from
\begin{equation}
\langle M_x^2\rangle=N[\,1+2\sum_{r\geq 1}C_x(r)\,]
\propto N\int_0^{+\infty}\dd r C_x(r)
\end{equation}
where changing the sum into integral is again allowed because of the diverging
characteristic length scale
$\kappa^{-1}$. Using now the scaling form (\ref{corrscaling}) we find that
\begin{equation}
\langle M_x^2\rangle\propto N \kappa^{-3/4}\propto N j_E^{-3/8} \, .
\label{mx2asymp}
\end{equation}
The above expression demonstrates the decrease of fluctuations with increasing
flux and it also tells us how the Gaussian distribution crosses over to the nontrivial
shape observed at the critical point.

\section{Final remarks}
\label{Final}

Returning to the problems discussed in the Introduction,
we can see that the connection between NESS and critical states
in terms of (universal) distribution functions is not straighforward.
NESS is generated by fluxes, and
fluxes may or may not generate long-range
correlations. There are numerous examples \cite{{SchmZia},{Marrobook},{Racz-LesH}}
where the fluxes are spatially
localized and long-range correlations do not develop (unless the system is at a
special point in the parameter space).
Clearly, in such cases, one cannot hope for a general description
to emerge. If the fluxes are global, as is the case for the model treated
in the present paper, long-range correlations do emerge frequently
\cite{{SchmZia},{Marrobook},{Racz-LesH}}.
Even in this case, however,
it is far from trivial whether these correlations
drive the system to an effectively critical state or whether they
make the system more rigid.

The driven transverse Ising model treated above is an example where
a global flux of energy generates long-range correlations but the resulting
state becomes more rigid in the sense that the fluctuations are decreased
due to the presence of the flux.
Driven diffusive systems \cite{SchmZia} provide other examples
\cite{derrida} where the
fluctuations decrease while the fluxes induce power-law correlations.
Thus we should conclude that, in general, the power-law correlations generated by
global fluxes cannot be the source of possible
universality of nonequilibrium distribution functions. It remains, however, an intriguing
question whether the weak long-range interactions supported by global fluxes
can underlie a kind of "weak" universality classification of distributions in NESS.


\section*{Acknowledgement}
We thank G. Gy\"orgyi, V. Hunyadi, and L. Sasv\'ari for helpful
discussions. This research has been partially supported by the Hungarian Academy
of Sciences (Grant No. OTKA T029792).

 
\end{document}